# Computational Investigation of Roughness Effects on Boundary-Layer Transition for Stetson's Blunt Cone at Mach 6

Arturo Rodriguez [1], Piyush Kumar [1], Cesar Diaz-Caraveo [1], Richard O. Adansi [1], Luis F. Rodriguez[2], and Vinod Kumar [3], *

1 Aerospace and Mechanical Engineering Department, University of Texas at El Paso, El Paso 79968, Texas
2 Mechanical and Aerospace Engineering, Clarkson University, Potsdam 13699, New York
3 Department of Mechanical and Industrial Engineering, Texas A&M University at Kingsville, Kingsville 78363, Texas
* Correspondence: vinod.kumar@tamuk.edu

**Abstract:** In this aerothermal study, we performed a two-dimensional steady-state Computational Fluid Dynamics (CFD) and heat conduction simulation at Mach 6. The key to our methodology was a one-way coupling between CFD surface temperature as a boundary condition and the calculation of the heat transfer flux and temperatures inside the solid stainless-steel body of a nose geometry. This approach allowed us to gain insight into surface heat transfer signatures with corresponding fluid flow regimes, such as the one experienced in laminar fluid flow. We have also examined this heat transfer under roughness values encountered in Stetson's studies at the Wright-Patterson Air Force Base Ludwig tube. To validate our findings, we have performed this type of work on a blunt cone, specifically for the U.S. Air Force. The research focuses on predicting transition onset using laminar correlations derived from Stetson's experimental studies, examining the role of discrete roughness elements. Findings emphasize the importance of incorporating non-equilibrium effects in future computational frameworks to enhance predictive accuracy for high-speed aerodynamic applications.

## 1. Introduction

Research has shown that understanding the transition from laminar to turbulent flow Boundary-Layer Transition (BLT) is essential in designing faster vehicles for defense and space applications. That is because such vehicles would experience lower heating levels at laminar flow, overcoming the limitations of high-speed temperature conditions[1]. BLT involves studying disturbances caused by geometric parameters and fluid flow conditions, such as surface roughness elements, acoustic noise, velocity shearing, and high-temperature, density, and pressure fluctuations. These disturbances ultimately lead to the development of turbulent spots and differential heating. Although past studies have used the Reynolds number, a dimensionless quantity, to predict turbulent flow within a system, the Reynolds number is widely recognized as unreliable as it cannot accurately predict the exact location of boundary layer transition. In some cases, experiments have revealed laminar flow at high Reynolds numbers and vice versa, indicating the potential for unexpected results in fluid dynamics.

The development of turbulent spots with a spanwise direction can be observed in flat surfaces, a phenomenon that is crucial to our understanding of fluid dynamics. These turbulent regions grow nonlinearly and get amplified into different types in modes of harmonicity[2]. Linear and Parabolic Stability Theory, a key development in BLT, was developed to predict transition[3]. By analyzing the conditions when the fluid flow parameters start to depart from linearity or have an amplified non-linear growth into turbulence, we can capture the transition amplification in the boundary layer leading to transition. However, amplification factors differ in geometry and conditions. When non-linear fluctuations occur and get broken into secondary instabilities, the fluid flow becomes chaotic, and non-linear anisotropic acceleration instances arise from the physics arising from conditions and geometric parameters. We can capture the transition of the boundary layer by performing physical experiments, but they are expensive, and physical configurations are limited. Despite many efforts by the community and some successes, no general computational solution to simulate different fluid flows and vehicle types that incorporate boundary layer transition currently exists.

This paper presents a novel approach to predicting the roughness-induced boundary layer transition for the blunt cone from Stetson's 1983 AIAA paper[4–6]. This transition, caused by a variety of mechanisms, is an exciting area of study that could depend on fluid shear with different velocities, Kelvin-Helmholtz instability, different densities between fluids, Rayleigh-Taylor instability, pressure fluctuations, Baroclinic instability, sound waves, swallowing of the entropy layer, roughness elements on the vehicle surface, concavity on the vehicle surface, Gortler instability, cross-flow streamlines colliding with vortices, and other mechanisms leading to the transition. We provide some graphical views of these instabilities[7–12].

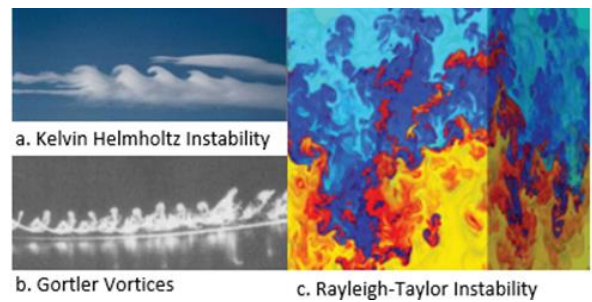

**Figure 1.** Transition Mechanisms. (a) Kelvin-Helmholtz Instability[9]. (b) Gortler Vortices[7]. (c) Rayleigh-Taylor Instability[8]

For this study, we have used computational tools to calculate the Reynolds moment thickness at the sonic point. This leads to a roughness value prediction based on the Stetson study, where the laminar correlations were created. To validate the transition range of Stetson to Mach 6 due to the roughness elements on the surface, we have used its Laminar correlations, commercial fluid flow solver, and commercial heat conduction solver to perform these predictions. The figures below are the geometry and domains for our numerical simulations. These simulations were used to calculate the roughness values of Stetson's 1983 Mach 6 blunt cone for transition to turbulence. Looking ahead, this study will pave the way for future validation and studies for a blunt ogive geometry at Holloman Air Force Base (HAFB), which we are excited to undertake[13–15].

In our research, we employed a commercial Navier-Stokes fluid flow solver for the fluid domain and a commercial heat conduction solver to solve the heat equation in the stainless-steel solid domain. The primary objective of this study is to predict the potential transition of the boundary layer at hypersonic speeds due to the presence of a roughness element(s) [16–18]. This investigation is motivated by the experimental work of Ken Stetson at the Ludwig tube in Wright-Patterson Air Force Base, where solid pebble particulates in the high enthalpy chamber were observed to move into the main channel of fluid flow in the shock tube. The impingement of these particulates onto the surface of the cone resulted in a roughness value of 50 micro-inches, ultimately leading to transition phenomena. We will use a laminar fluid flow solution to the Navier-Stokes Equations for our prediction. Furthermore, we intend to provide a visual representation of the laminar solution of the fluid flow alongside Ken Stetson's Schlieren shadowgraph image.

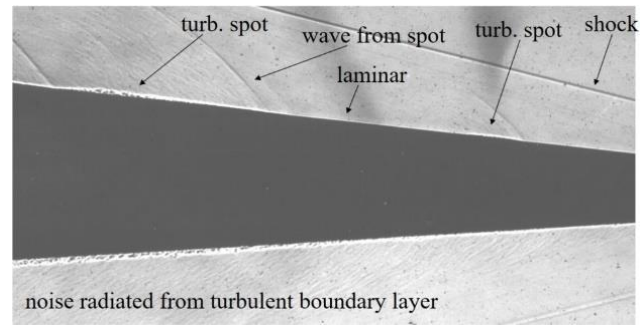

**Figure 2.** Stetson's Blunt Cone | Mach 6 Test | Schlieren Imaging[6]

The Stetson Schlieren image, a visual representation, reveals the intricate formation of shock waves and the transition of boundary layers through turbulent points and Fedorov transition phases. It also uncovers the sound radiated from the bottom surface[19]. The main transition mechanisms of this cone, as depicted in the image, are the roughness elements caused by the particles colliding with the fluid flow and hitting the vehicle surface, generating roughness elements. Furthermore, the image below showcases the crossflow instability, a phenomenon that combines Gortler vortices and streamline impact[12].

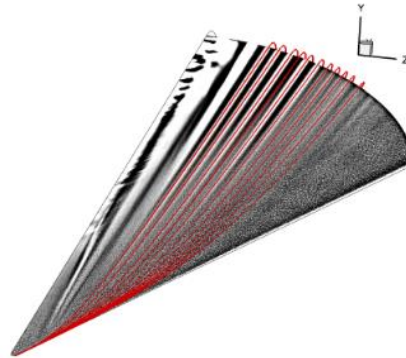

**Figure 3.** Crossflow Instability [12]

The U.S. Department of Defense has displayed significant interest in hypersonic vehicles owing to their potential tactical and strategic application within Intercontinental Ballistic Missiles (IBMs), Unmanned Aerial Vehicles (UAVs), and scramjets. This interest has consequently driven advancements in high-speed flight geometries[20]. Initial design approaches featured sharp and blunt noses, with sharp noses offering minimal drag but susceptibility to ablation. Subsequent shifts in design philosophy saw the adoption of ogival geometries for reentry vehicles, primarily to enhance vehicle survivability[21]. This shift resulted in the propagation of shock waves and more uniformly distributed heating. The management and reduction of fluid flow have proven pivotal in optimizing the design for ablation and in developing extreme-temperature materials such as ultra-high-temperature ceramics (UHTCs) and carbon-based materials[22,23]. These advancements reflect the continued evolution of high-speed flight vehicles, underscoring the need for continued research and development in this domain.

Hypersonic vehicles encounter a challenging environment in hypersonic flight with high aero-thermal heating loads and skin friction. The fluid flow around these vehicles creates harsh conditions, making transition prediction and control critical factors in achieving higher speeds and serving as vital considerations in warfare. Stakeholders must emphasize the importance of these factors as such. Nevertheless, there is optimism regarding the potential to comprehend and optimize vehicle geometry, which can delay the transition to turbulence and reduce the cost of material requirements for heat shields. Aeronautics, operating in various flow regimes, each with its unique impact on vehicle performance, presents a promising avenue for overcoming these challenges[24].

**2. Methods**

We employed a compressible steady-state viscous Navier-Stokes solver to address this problem[25,26]. We solve the following governing equations:

Conservation of mass, momentum, and energy:

$$\frac{\partial}{\partial x_j}(\rho u_j) = 0 \tag{1}$$

$$\frac{\partial}{\partial x_j}(\rho u_i u_j) = -\frac{\partial p}{\partial x_i} + \frac{\partial}{\partial x_j}(\tau_{ij}), i = 1,2 \tag{2}$$

$$\frac{\partial}{\partial t}\left(\rho e + \frac{1}{2}\rho u^2\right) + \frac{\partial}{\partial x_j}\left[\left(\rho e + \frac{1}{2}\rho u^2\right)v_j\right] = -\frac{\partial}{\partial x_j}(pu_j) + \frac{\partial}{\partial x_j}(\tau_{ij}u_i) + \frac{\partial}{\partial x_j}(\dot{q}_j) \tag{3}$$

In our research, we have employed a particular equation formulation, informed by physical experiments, to approximate the desired solution. To accurately depict hypersonic physics, we must adjust the energy equation to encompass classical and quantum effects within the chemical non-equilibrium, such as the Two-Temperature Model. Employing the finite volume spectral method, we have utilized this formulation with an implicit steady-state solver and the AUSM flux type to solve the problem. It should be noted that this approach only encompasses the physics of the fluid domain[27].

We also solve the heat equation for the heat conduction domain:

$$-\nabla \cdot k\nabla T = 0 \tag{4}$$

In thermal analysis, the surface boundary condition primarily concerns the solver's transient aspect and pertains solely to spatial considerations. The finite element method effectively addresses the Partial Differential Equation (PDE) associated with this scenario.

In this study, the Navier-Stokes solver was coupled with a heat conduction solver to enable direct comparison with experimental data. The primary variable for comparison is the heat transfer coefficient. This process involves using the surface temperature obtained from the Navier-Stokes solver as a boundary condition in the heat conduction solver for the solid domain. The heat transfer coefficient is then calculated based on these results and compared against the experimental measurements from Stetson's work. This coupling provides a consistent framework for validating numerical simulations against physical experimental data.

The laminar correlations described in Schneider's 2004 review paper provide critical tools for predicting and locating boundary-layer transitions on high-speed vehicles due to surface roughness. These correlations enable the determination of transition onset and location along the vehicle's surface, given the roughness height and the flow conditions experienced during testing. By relating parameters such as surface roughness, temperature, Reynolds number, and momentum thickness, these correlations allow precise predictions for the transition region, typically confined to the nose and frustum sections of the vehicle. The following figures, adapted from Schneider's Aerospace Sciences Review, illustrate these laminar correlations, emphasizing their application to blunt cones in hypersonic flow regimes.

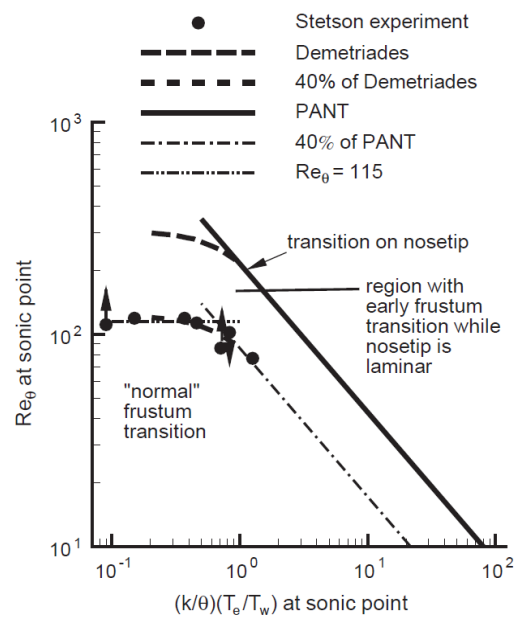

**Figure 4.** Roughness-Induced Transition Laminar Correlations [6]

### 3. Computational Set-Up and Results

In our study, we meticulously constructed a computational model by employing a Navier-Stokes solver to simulate the fluid flow around the blunt cone and utilized a heat equation commercial solver to model heat conduction. The blue section represents the fluid flow domain, while the dark gray depicts the heat conduction domain in Figure 4. The dimensions of the domains are as follows: The base radius is 2.0 inches, and the nose tip base has a measurement of 0.6 inches with a half angle of 8 degrees. These specific dimensions pertain to the blunt cone utilized by Stetson in his physical experiments at the Wright-Patterson Air Force Base.

The purpose of using this cone was to observe how roughness elements can induce transition due to discontinuities on the surface, thus causing disturbances in fluid flow. The varying heights of the roughness elements lead to diverse fluid flow development and the formation of distinct environments, which have practical implications for understanding and predicting fluid flow in real-world scenarios[28–32]. As a consequence of these discontinuities, separation occurs, triggering the transition of the boundary layer. This process results in the formation of horseshoe vortices that are pushed upstream as the fluid flow stagnates, ultimately creating signatures of small shock waves during the development of the fluid flow. Thus, it represents an unsteady time of the simulation, a significant finding with practical implications in fluid dynamics and aerodynamics.

The remarkable aspect of Computational Fluid Dynamics (CFD) and its utility lies in its capability to non-intrusively measure fluid flow phenomena without causing disruptions. Unlike traditional methods, CFD does not introduce obstructions that alter the fluid flow configuration, allowing for the validation of scenarios at precise locations by using thermocouples and pressure sensors located within the interior of the blunt cone. This seamless amalgamation of physical experiments with numerical simulations is a testament to CFD's effectiveness.

For this study, which mimicked the Stetson Mach 6 physical test, we initialized the fluid flow at Mach 6 to reach a steady state. This simulation is an axisymmetric simulation around the z-axis. We have initialized the inlet and outlet conditions for the surrounding conditions at 4000 ft altitude. Static Pressure being 1827.71 lbs/ft² and Static Temperature being 504.1 R. The boundary condition at the vehicle wall is a no-slip boundary condition. We consider a stainless steel 17-4 PH wall with variable conditions for the heat conduction domain. The conductivity and specific heat are functions of temperature,

$$k = 2.08e - 4 + 1.13e - 7 \cdot T \tag{5}$$
$$C_p = 0.104 + 3.38e - 5 \cdot T + 4.45e - 8 \cdot T^2 \tag{6}$$

Density ($\rho$) is 0.282 Lbm/in³ and we have an inner adiabatic wall. This means having a zero flux Neumann boundary condition. The wall thickness on the z-axis is a unit of 1 in.

The current study applies boundary conditions to a computational fluid dynamics (CFD) simulation of a compressible viscous fluid flow domain with an axisymmetric configuration. Specific pressure far-fields are applied to the front boundaries to represent compressible fluid flows. Additionally, the simulation utilizes the ideal gas condition for the density fluid flow as an atmospheric pressure outlet at an altitude of 4000 ft. Notably, the cone front surface features crucial no-slip stationary walls.

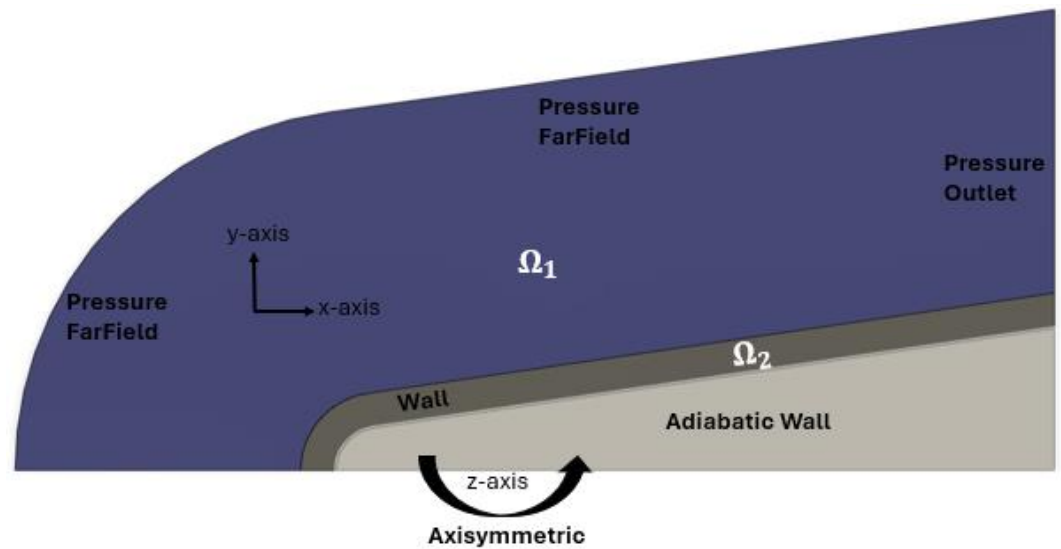

**Figure 5.** Computational Set-Up

In our simulation, the mesh exhibits low skewness, high orthogonality, and high element quality for non-deformed orthogonal structured elements. Maintaining an aspect ratio of four or less is crucial for accurate numerical convergence and stability. We focus on the vehicle noses, the section with the most effect on the fluid flow.

In our research, using drag force as a critical parameter for achieving mesh independence holds considerable significance. Our study analyzed four different element sizes: mesh 1, coarse; mesh 2, base; mesh 3, fine; and mesh 4, ultra-fine mesh sizes, as illustrated in the accompanying table. Upon stabilizing the drag force, it was evident that we noticed convergence with the element size in the case under consideration.

**Table 1.** Mesh Independence Study

| Mesh | Element Size | Elements | Steady-State Drag Force |
|---|---|---|---|
| 1 | $2 \times 10^{-3}$ | 11821 | 1110 |
| 2 | $1 \times 10^{-3}$ | 47111 | 1127 |
| 3 | $6 \times 10^{-4}$ | 130534 | 1136 |
| 4 | $4 \times 10^{-4}$ | 293138 | 1144 |

The properties represent the behavior of air as an ideal gas, assuming standard atmospheric conditions at an altitude of 4000 ft. The formulation employs an implicit approach with the AUSM flux type, a key technical detail, utilizing a least squares cell-based gradient for spatial discretization and a second-order upwind scheme for flow. This comprehensive methodology facilitates a thorough analysis. A Courant number of 5 has been utilized, and the resultant output is a heat transfer boundary condition that will be applied to a 17-4 PH stainless steel material. The outcomes include contour results for a blunt cone geometry.

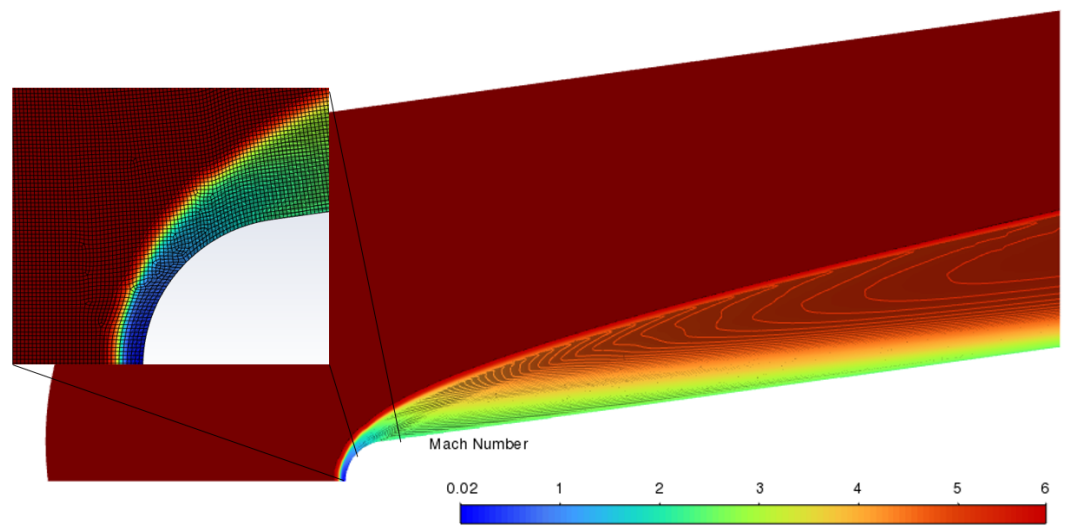

**Figure 6.** Blunt Cone | Laminar Solution | Mach Number Contour

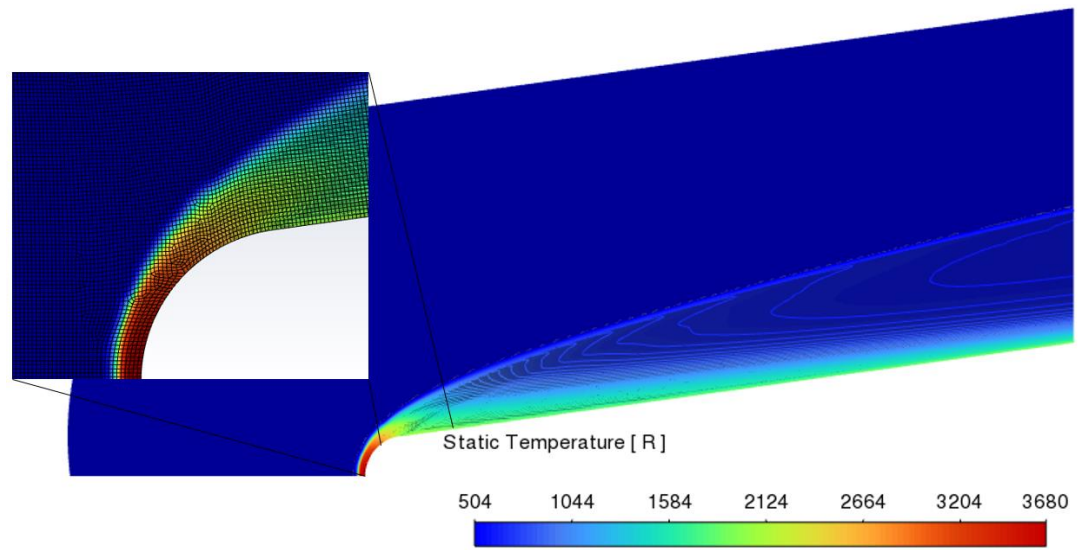

**Figure 7.** Blunt Cone | Laminar Solution | Temperature Contour

In heat conduction, it is essential to establish clear boundary conditions in the finite element method. This method involves utilizing basis functions between nodes to approximate the solution. We approached the problem using the weak form of the partial differential equation, which represents the variational integral form of the equation. The initial input depends on the boundary constant temperature conditions obtained through computational fluid dynamics (CFD) simulations, while we set other conditions as axisymmetric with zero Neumann boundary constraints.

In our investigation, we integrated the fluid flow and the Navier-Stokes solver with heat conduction, specifically the Heat Equation solver. By doing so, we could derive the aerothermal results, which were then quantified using the heat transfer coefficient. Our computational analysis yielded results relevant to the physical experiments outlined in Stetson's seminal 1983 paper, presented at the Fluid and Plasma AIAA conference[4].

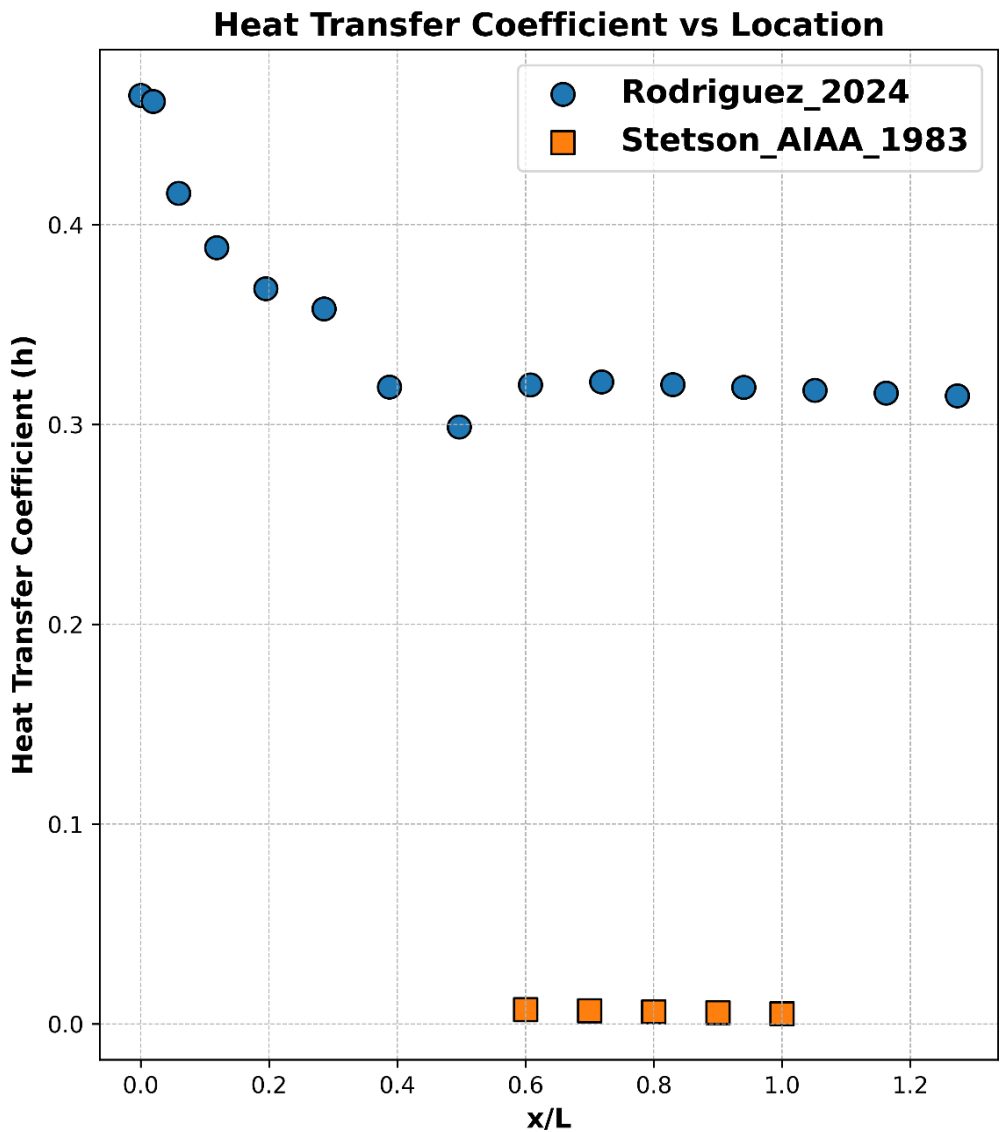

**Figure 8.** Blunt Cone at Mach 6 | Laminar Fluid Flow

The observed discrepancy in the heat transfer coefficient results between Rodriguez_2024 and Stetson_AIAA_1983 can be attributed to the absence of a more advanced energy equation model, such as the two-temperature model. As highlighted in prior literature, including Schneider's review, accurate modeling of high-enthalpy hypersonic flows requires energy equations that account for non-equilibrium effects[6]. These effects, which include vibrational relaxation and dissociation processes, play a significant role in the complex energy exchange mechanisms. Without incorporating these phenomena, the simulations lack the fidelity to fully resolve the aerothermal behavior in extreme flow conditions.

Stetson's experiments highlight the critical role of shock-induced vibrational excitation in shaping boundary-layer development and subsequent heat transfer behavior. Similarly, Schneider's review underscores that discrepancies between predicted and experimental results frequently stem from inadequate modeling of non-equilibrium processes. The absence of a two-temperature framework in the current computations limits their ability to account for key mechanisms, such as the interaction between vibrational-electronic energy modes and dissociation processes, significantly influencing boundary-layer evolution and instability growth.

Future work will address these limitations by incorporating a two-temperature model to resolve vibrational-electronic energy exchanges and dissociation coupling terms. These enhancements are expected to improve the alignment of computational results with experimental observations by capturing the intricate interplay between relaxation processes and boundary-layer instabilities. As Schneider emphasizes, accurately modeling these phenomena is essential to advancing the predictive fidelity of aerothermal simulations in high-enthalpy hypersonic flows.

**4. Discussion**

The investigation of boundary-layer transition over hypersonic cones has consistently underscored the sensitivity of laminar-turbulent transition to surface conditions, particularly roughness height. For the Stetson cone at Mach 6, roughness-induced instabilities dominate when surface perturbations are introduced at a scale of 60 micro inches. Such conditions are absorbing for experimental and computational studies to resolve transition mechanisms.

Experimental studies highlight that introducing discrete roughness elements, such as those with a height of 60 micro inches, effectively triggers localized disturbances within the boundary layer. These disturbances evolve through linear amplification mechanisms into fully turbulent flow downstream. This finding aligns with the framework of Mack's instability theory, where second-mode instabilities are particularly susceptible to small perturbations under hypersonic flow conditions.

Schneider's review underscores that the amplification of roughness-induced instabilities depends not only on the height of the roughness but also on its streamwise position relative to the entropy layer swallowing length [6]. The transition onset shifts upstream as the interaction of roughness height with the boundary-layer shear leads to an earlier breakdown of laminar flow. This behavior mirrors the trend observed in Stetson's experimental data, where surface roughness significantly modulates the transition Reynolds number.

**5. Conclusion**

The present study demonstrates the role of surface roughness on the boundary-layer transition for a blunt cone geometry at Mach 6, providing computational predictions based on laminar correlations derived from Stetson's seminal experiments. These results underscore the sensitivity of boundary-layer stability to discrete roughness elements and validate the robustness of Navier-Stokes solvers in capturing heat transfer and instability phenomena.

While the laminar correlations predict transition locations with reasonable accuracy, discrepancies in heat transfer coefficients, as compared to Stetson's experimental data, highlight the limitations of the current approach. These differences can largely be attributed to the absence of thermochemical nonequilibrium modeling, such as the two-temperature framework, which is crucial for capturing the interaction between vibrational relaxation and dissociation effects in high-enthalpy flows.

Future work should integrate advanced energy equation models to address these gaps, including vibrational and electronic energy mode contributions. As emphasized in Schneider's review, such enhancements are essential for improving predictive fidelity and aligning computational results more closely with experimental observations. The coupling of high-fidelity computational tools with carefully controlled experiments will advance the understanding of hypersonic transition phenomena and support the development of reliable design methodologies for high-speed vehicles.

These findings reinforce the need for continued refinement in computational methodologies and experimental validation, aligning with the broader goals of hypersonic aerothermal research. By addressing the limitations of current models and incorporating additional physical effects, the predictive capability for transition and heat transfer phenomena can be significantly enhanced.

**Author Contributions:** Conceptualization, V.K., and A.R.; methodology, A.R., C.D.C.; software, A.R., and P.K.; investigation, A.R., C.D.C.; resources, A.R.; writing—original draft preparation, A.R.; writing—review and editing, C.D.C., L.F.R., and R.O.A.; visualization, R.O.A..; supervision, V.K., and P.K.; project administration, A.R., and C.D.C.; funding acquisition, V.K. All authors have read and agreed to the published version of the manuscript.

**Funding:** The National Science Foundation Graduate Research Fellowship Program (NSF GRFP) under grant number 226101161A. The Air Force Office of Scientific Research funded this research under the Agile Science of Test and Evaluation (T&E) program grant number FA9550-19-1-0304. The U.S. Department of Energy Minority Serving Institutions Partnership Program (DOE-MSIPP) funded this research under grant number GRANT13584020.

**Data Availability Statement:** The data presented in this study are available from the corresponding author upon reasonable request.

**Acknowledgments:** I acknowledge Prof. Steven P. Schneider, Robert P. Velte, and Owen States from the School of Aeronautics and Astronautics at Purdue University for their guidance in this work.

**Conflicts of Interest:** The authors declare no conflicts of interest.

**Nomenclature:**

| | | | | | |
|---|---|---|---|---|---|
| $\rho$ | Density | $\left(\frac{Lbm}{in^3}\right)$ | $x$ | Spatial Dimension | $in$ |
| $u$ | Velocity | $\left(\frac{in}{s}\right)$ | $p$ | Pressure | $\left(\frac{lbs}{ft^2}\right)$ |
| $\tau$ | Shear Stress | $\left(\frac{lb}{in^2}\right)$ | $t$ | Time | $s$ |
| $e$ | Energy | $BTU$ | $\dot{q}$ | Heat Transfer Rate | $\left(\frac{BTU}{s}\right)$ |
| $T$ | Temperature | $R$ | $k$ | Conductivity | $\left(\frac{BTU}{in \cdot s \cdot F}\right)$ |
| $Cp$ | Heat Capacity | $\left(\frac{BTU}{Lbm \cdot F}\right)$ | $h$ | Heat Transfer Coefficient | $\left(\frac{BTU}{ft^2 \cdot s \cdot R}\right)$ |
| $F$ | Drag Force | $lbf$ | | | |

**References**

1. Caraveo, C.D.; Rodriguez, A.; Kumar, V.; Muñoz, J.A.; Wolk, K.; Miesner, S.; Montemayor, M.; Daimaru, T.; Furst, B.I.; Roberts, S.N. Performance Dryout Limits of Oscillating Heat Pipes: A Comprehensive Theoretical Prediction and Experimental Determination. *AIAA SciTech Forum and Exposition, 2024* **2024**, doi:10.2514/6.2024-0657.

2. Casper, K.M.; Beresh, S.J.; Schneider, S.P. Pressure Fluctuations beneath Instability Wavepackets and Turbulent Spots in a Hypersonic Boundary Layer. *J Fluid Mech* **2014**, *756*, 1058–1091, doi:10.1017/jfm.2014.475.

3. Paredes, P. Advances in Global Instability Computations: From Incompressible to Hypersonic Flow. *Doctoral Thesis* **2014**.

4. STETSON, K. Nosetip Bluntness Effects on Cone Frustum Boundary Layer Transition in Hypersonic Flow. **1983**, doi:10.2514/6.1983-1763.

5. Jewell, J.S.; Kimmel, R.L. Boundary-Layer Stability Analysis for Stetson's Mach 6 Blunt-Cone Experiments. *J Spacecr Rockets* **2017**, *54*, 258–265, doi:10.2514/1.A33619.

6. Schneider, S.P. Hypersonic Laminar-Turbulent Transition on Circular Cones and Scramjet Forebodies. *Progress in Aerospace Sciences* **2004**, *40*, 1–50, doi:10.1016/j.paerosci.2003.11.001.

7. Floryan, J.M. On the Görtler Instability of Boundary Layers. *Progress in Aerospace Sciences* **1991**, *28*, 235–271, doi:10.1016/0376-0421(91)90006-P.

8. Boffetta, G.; Mazzino, A. Incompressible Rayleigh-Taylor Turbulence. *Annu Rev Fluid Mech* **2017**, *49*, 119–143, doi:10.1146/annurev-fluid-010816-060111.

9. Smyth, W.D.; Moum, J.N. Ocean Mixing by Kelvin-Helmholtz Instability. *Oceanography* **2012**, *25*, 140–149, doi:10.5670/oceanog.2012.49.

10. Wagnild, R.M. High Enthalpy Effects on Two Boundary Layer Disturbances in Supersonic and Hypersonic Flow. *GPPS Chania20* **2012**, *Ph.D.*

11. Siddiqui, F.; Gragston, M.; Saric, W.S.; Bowersox, R.D.W. Mack-Mode Instabilities on a Cooled Flared Cone with Discrete Roughness Elements at Mach 6. *Exp Fluids* **2021**, *62*, doi:10.1007/s00348-021-03304-6.

12. Dinzl, D.J.; Candler, G. V. Direct Simulation of Hypersonic Crossflow Instability on an Elliptic Cone. *AIAA Journal* **2017**, *55*, 1769–1782, doi:10.2514/1.J055130.

13. Terrazas, J.A. Computational Analysis of Water Braking Phenomena for High-Speed Sled and Its Machine Learning Framework.

14. Pérez, J.; Baez, R.; Terrazas, J.; Rodríguez, A.; Villanueva, D.; Fuentes, O.; Kumar, V.; Paez, B.; Cruz, A. Physics-Informed Long-Short Term Memory Neural Network Performance on Holloman High-Speed Test Track Sled Study. *American Society of Mechanical Engineers, Fluids Engineering Division (Publication) FEDSM* **2022**, *2*, doi:10.1115/FEDSM2022-86953.

15. Terrazas, J.; Rodriguez, A.; Kumar, V.; Adansi, R.; Krushnarao Kotteda, V.M. Three-Dimensional Two-Phase Flow Simulations of Water Braking Phenomena for High-Speed Test Track Sled. *American Society of Mechanical Engineers, Fluids Engineering Division (Publication) FEDSM* **2021**, *1*, doi:10.1115/FEDSM2021-65799.

16. Wheaton, B.M.; Bartkowicz, M.D.; Subbareddy, P.K.; Schneider, S.P.; Candler, G. V. Roughness-Induced Instabilities at Mach 6: A Combined Numerical and Experimental Study. *41st AIAA Fluid Dynamics Conference and Exhibit* **2011**, doi:10.2514/6.2011-3248.

17. De Tullio, N.; Paredes, P.; Sandham, N.D.; Theofilis, V. Laminar-Turbulent Transition Induced by a Discrete Roughness Element in a Supersonic Boundary Layer. *J Fluid Mech* **2013**, *735*, 613–646, doi:10.1017/jfm.2013.520.

18. Schneider, S.P. Effects of Roughness on Hypersonic Boundary-Layer Transition. *J Spacecr Rockets* **2008**, *45*, 193–209, doi:10.2514/1.29713.
19. Fedorov, A. Transition and Stability of High-Speed Boundary Layers. *Annu Rev Fluid Mech* **2011**, *43*, 79–95, doi:10.1146/annurev-fluid-122109-160750.
20. Urzay, J. Supersonic Combustion in Air-Breathing Propulsion Systems for Hypersonic Flight. *Annu Rev Fluid Mech* **2018**, *50*, 593–627, doi:10.1146/annurev-fluid-122316-045217.
21. Amar, A.J.; Cooper, J.; Oliver, A.B.; Salazar, G.; Agricola, L. Mesh Deformation Boundary Conditions for Three-Dimensional Ablation Solvers. *AIAA Science and Technology Forum and Exposition, AIAA SciTech Forum 2022* **2022**, doi:10.2514/6.2022-1642.
22. Cedillos-Barraza, O.; Manara, D.; Boboridis, K.; Watkins, T.; Grasso, S.; Jayaseelan, D.D.; Konings, R.J.M.; Reece, M.J.; Lee, W.E. Investigating the Highest Melting Temperature Materials: A Laser Melting Study of the TaC-HfC System. *Sci Rep* **2016**, *6*, doi:10.1038/srep37962.
23. Cedillos-Barraza, O.; Grasso, S.; Nasiri, N. Al; Jayaseelan, D.D.; Reece, M.J.; Lee, W.E. Sintering Behaviour, Solid Solution Formation and Characterisation of TaC, HfC and TaC-HfC Fabricated by Spark Plasma Sintering. *J Eur Ceram Soc* **2016**, *36*, 1539–1548, doi:10.1016/j.jeurceramsoc.2016.02.009.
24. Paredes, P.; Choudhari, M.M.; Li, F.; Jewell, J.S.; Kimmel, R.L.; Marineau, E.C.; Grossir, G. Nose-Tip Bluntness Effects on Transition at Hypersonic Speeds. *J Spacecr Rockets* **2019**, *56*, 369–387, doi:10.2514/1.A34277.
25. Versteeg, H.K.; Malalasekera, W. The Finite Volume Method. *An Introduction to Computational Fluid Dynamics* **1995**, 102–206.
26. Sørensen, L.S. An Introduction to Computational Fluid Dynamics: The Finite Volume Method. **1999**, *M*, 267.
27. Candler, G. V.; Subbareddy, P.K.; Brock, J.M. Advances in Computational Fluid Dynamics Methods for Hypersonic Flows. *J Spacecr Rockets* **2015**, *52*, 17–28, doi:10.2514/1.A33023.
28. Rodriguez, A.; Enriquez, A.; Terrazas, J.; Villanueva, D.; Paez, B.; Dudu, N.; Baez, R.; Harris, C.; Kumar, V. Predicting Boundary-Layer Transition (BLT) Using Artificial Intelligence (AI) Causality Inference. *APS March Meeting Abstracts* **2022**, *2022*, S49-005.
29. Rodriguez, A.; Terrazas, J.; Adansi, R.; Krushnarao Kotteda, V.M.; Munoz, J.A.; Kumar, V. Causal Inference Analysis to Find Relationships Found in Boundarylayer Transition Part i: Theoretical. *American Society of Mechanical Engineers, Fluids Engineering Division (Publication) FEDSM* **2021**, *1*, doi:10.1115/FEDSM2021-61843.
30. Sendstad, O. *The Near-Wall Mechanics of Three-Dimensional Turbulent Boundary Layers*; Stanford University, 1992; ISBN 9798208712337.
31. White, F.M.; Majdalani, J. *Viscous Fluid Flow*; McGraw-Hill New York, 2006; Vol. 3;.
32. Wu, Z.; Zaki, T.A.; Meneveau, C. High-Reynolds-Number Fractal Signature of Nascent Turbulence during Transition. *Proc Natl Acad Sci U S A* **2020**, *117*, 3461–3468, doi:10.1073/pnas.1916636117.